\documentclass[a4paper, amsfonts, amssymb, amsmath, reprint, showkeys, nofootinbib, twoside]{revtex4-1}

\pdfoutput=1

\usepackage[compat=1.0.0]{tikz-feynman}
\usepackage{subfigure}

\usepackage[english]{babel}
\usepackage[utf8]{inputenc}
\usepackage[colorinlistoftodos, color=green!40, prependcaption]{todonotes}
\usepackage[pdftex, pdftitle={Article}, pdfauthor={Author}]{hyperref} 
\begin{document}
\title{Observable features  in (ultra)high energy neutrinos due to active-sterile secret interactions}

\author{Damiano Fiorillo, Gennaro Miele, Stefano Morisi}
 %   \email[Correspondence email address: ]{email@institution.com}% Your name
    \affiliation{Dipartimento di Fisica {\it "Ettore Pancini"}, Universit\`a degli studi di Napoli Federico II, Complesso Univ. Monte S. Angelo, I-80126 Napoli, Italy}
    \affiliation{INFN - Sezione di Napoli, Complesso Univ. Monte S. Angelo, I-80126 Napoli, Italy}
    
\author{Ninetta Saviano}
	\affiliation{INFN - Sezione di Napoli, Complesso Univ. Monte S. Angelo, I-80126 Napoli, Italy}

%\author{Damiano Fiorillo, Gennaro Miele, Stefano Morisi}
  %  \affiliation{Dipartimento di Fisica {\it "Ettore Pancini"}, Universit\`a degli studi di Napoli Federico II, Complesso Univ. Monte S. Angelo, I-80126 Napoli, Italy}
   % \affiliation{INFN - Sezione di Napoli, Complesso Univ. Monte S. Angelo, I-80126 Napoli, Italy}
    
%\author{Ninetta Saviano}
%	\affiliation{INFN - Sezione di Napoli, Complesso Univ. Monte S. Angelo, I-80126 Napoli, Italy}

\date{\today} % Leave empty to omit a date

\begin{abstract}
We consider the effects of active-sterile secret neutrino interactions, mediated by a new pseudoscalar particle,  on high- and  ultra high-energy neutrino fluxes. In particular, we focus on the case of 3 active and 1 sterile neutrino coupled by a  flavor dependent  interaction, extending the case of 1 active  and 1 sterile neutrino we have recently examined. We find that, depending on the kind of interaction of sterile neutrino with  the active sector,  new regions of the parameter space for secret interactions  are  now allowed, with the masses of sterile neutrino and scalar mediator ranging from 10 MeV to 1 GeV, leading to interesting  phenomenological implications on  two benchmark fluxes we consider, namely  an astrophysical power law flux,  in the range below 100 PeV, and a cosmogenic flux, in the Ultrahigh energy range. First of all, the final  active fluxes can present a measurable depletion observable in future experiments.  Especially, in the case of only $\nu_\tau-\nu_s $ interaction, we find that the effects on the astrophysical power law flux can be so large to be already  probed by the IceCube experiment.  Moreover,  we find  intriguing  features in the energy dependence of the flavor ratio.

\end{abstract}

\maketitle

\section{Introduction}

High energy neutrinos can be produced by the interactions of High and Ultrahigh Energy Cosmic rays. 
Photohadronic \cite{Winter:2013cla,Murase:2015xka} and hadron-hadron \cite{Loeb:2006tw,Murase:2013rfa,Tamborra:2014xia,Bechtol:2015uqb} interactions of High Energy cosmic rays in astrophysical objects, such as Active Galactic Nuclei (AGN), Gamma-Ray Bursts (GRB), Starburst Galaxies (SBG), can produce high energy neutrinos with energy up to the order of 100 PeV.
On the other hand, photohadronic interaction  of Ultrahigh Energy Cosmic rays  (UHECRs) with the photons of the Cosmic Microwave Background (CMB) \cite{Beresinsky:1969qj} can give rise to cosmogenic neutrinos \cite{Berezinsky:1998ft,Nagano:2000ve,Engel:2001hd,Kusenko:2001gj,Anchordoqui:2002hs,Fodor:2002hy,Kalashev:2002kx,Semikoz:2003wv,Fodor:2003ph,Ave:2004uj,Seckel:2005cm,DeMarco:2005kt,Allard:2006mv,Becker:2007sv,Anchordoqui:2007fi,Berezinsky:2010xa,Ahlers:2010fw,Katz:2011ke,Gelmini:2011kg,Ng:2014pca,Murase:2014tsa,Heinze:2015hhp,Aloisio:2015ega,Halzen:2016gng,Cherry:2018rxj,Vitagliano:2019yzm} with energies in the range  $(10^2\div 10^{10})$PeV.

Recently, in  \cite{Fiorillo:2020jvy}  we have studied the possibility that the cosmogenic neutrino flux suffers from a measurable depletion (called absorption effect) observable in future experiments, due to the presence of active-sterile secret neutrino interactions. In particular,  in the  scheme of 1 active neutrino and 1 sterile, we have shown that the absorption effect is maximal for energies around $10^{9\div 10}$GeV, and it could be observed at  experiments like GRAND \cite{Alvarez-Muniz:2018bhp} and ARIANNA \cite{Anker:2019rzo}.

While active-active secret interactions have been thoroughly investigated in the literature \cite{Davidson:2003ha,Antusch:2008tz,Miranda:2004nb,Fornengo:2001pm,Huber:2001zw,Barranco:2005ps,Farzan:2017xzy,Ribeiro:2007ud,Coloma:2015kiu,deGouvea:2015ndi,Forero:2011pc,Mangano:2006ar,Ng:2014pca,Ioka:2014kca,Bakhti:2018avv,Kolb:1987qy, Archidiacono:2013dua,Laha:2013xua,Forastieri:2019cuf,Bustamante:2020mep,Blum:2014ewa,Murase:2019xqi,Babu:2019iml,Bustamante:2020mep}, active-sterile secret interactions are still under investigation. Previous studies \cite{Shoemaker:2015qul,Fiorillo:2020jvy} have analyzed some possible effects on the astrophysical fluxes, but a complete analysis of the constraints, arising in this model from cosmology, astrophysics and particle physics, is still lacking.

In this paper we approach this issue, investigating the more general case $3\; \& \; 1$  (3 active and 1 sterile neutrino $\nu_s$), where the interaction is possibly flavor dependent and mediated by a pseudoscalar particle $\varphi$.
In comparison with our previous work, we examine in detail the constraints on the parameter space of the secret interaction, coming both from laboratory experiments and from cosmological observations. We find that new regions  for the parameters are allowed  compared to the safe one we had investigated in the previous paper.
Extending the analysis to these new regions, we find again an absorption feature in the active neutrino fluxes, which can however occur over a much wider energy scale, reaching energies as low as $10^{6\div 7}$ GeV. Therefore, we consider two benchmark fluxes: an astrophysical power law flux, meant as a representative of the cumulative neutrinos from astrophysical objects, in the range below $100$ PeV, and a cosmogenic flux, in the Ultrahigh energy range. We find that the effects on the astrophysical power law flux can be so important to be probed by the IceCube experiment \cite{Aartsen:2013jdh}. More specifically, the effect of the interaction can produce a cutoff-like feature in the spectrum, which could potentially be an explanation for the lack of observed events above $10$ PeV.
Finally, we use the full  $3\; \& \; 1$  framework to generalize our predictions to the flavor structure modifications induced by the new interaction, finding interesting features in the energy dependence of the flavor ratio.  This is especially important from a phenomenological point of view  since flavor identification is performed differently in various energy ranges. In the IceCube energy range, flavor distinction can be partially achieved using the different topologies of cascade and track events. Further, tau neutrinos can give rise to a unique topology, the so called double bang events. Finally, electron antineutrinos should produce as a signature a peak in their interaction with the detector, due to the Glashow resonance in the process mediated by the $W^-$ boson. In the ultrahigh energy range flavor identification is more complicated: it can be partially achieved by distinguishing between the air showers produced by neutrinos of all flavors in the atmosphere, and the showers produced by skimming tau leptons. The latter originate from tau neutrinos passing through the Earth. This distinction therefore would in principle allow an identification of the tau neutrino ratio in the ultrahigh energy range. We therefore reach the conclusion that a combined analysis of the energy and flavor structure of the astrophysical neutrino fluxes in the energy region above the PeV would allow to provide definite informations about the possibility of active-neutrino sterile interactions.

The outline of the paper is the following: in Section \ref{sec:model} we specify our model, emphasizing the role of the parameters. We then discuss in Section \ref{sec:constraints}  the constraints, coming from laboratory experiments, cosmological observations and astrophysical data. A description of the benchmark fluxes we have analyzed, as well as of the methods used to describe the effects of the interaction, is provided in Section \ref{sec:fluxes}. In Section \ref{sec:results} we show our results and discuss it. Finally, in Section \ref{sec:conclusions}, we come to our conclusions.

\vskip5.mm

\section{Model} \label{sec:model}
In this section we  describe the model of active-sterile neutrino interaction analyzed for this work. For definiteness we  assume throughout the paper that neutrinos are described by Majorana spinors. We also consider just one sterile neutrino $\nu_s$ coupling with the active ones {\it via} a new interaction given by
\begin{eqnarray}
\mathcal{L}_{\rm{SI}} = \sum_\alpha \lambda_\alpha \, \overline{\nu}_\alpha \gamma_5 \nu_s \varphi \,,\label{coupling}
\end{eqnarray}
where $\alpha=e,\mu,\tau$ and $\lambda_\alpha$ are  dimensionless free couplings. We have to assume $\lambda$ imaginary, since, while the scalar contraction $\overline{\nu}_\alpha \nu_s$ is purely real for Majorana spinors, the pseudoscalar combination $\overline{\nu}_\alpha \gamma_5 \nu_s$ is instead purely imaginary. For clarity we have chosen a model with pseudoscalar interaction and imaginary coupling constant. However, it should be noted that, for energies much larger than the active neutrino masses, all of our results are unchanged in the model with real coupling and scalar interaction. In order to maintain parity, only one of the two interactions can be allowed, and they cannot appear simultaneously: consequently, in the following the mediator of the interaction $\varphi$ is chosen to be a pseudoscalar.

The interaction in eq.(\ref{coupling}) is assumed to arise after the breaking of $SU_L(2)$ weak group, since it explicitly violates it. The study of a  complete Standard Model Lagrangian is beyond the scope of the present paper, since we are only interested into the phenomenological consequences of the interaction  (\ref{coupling}).
Nevertheless, it is interesting to observe that our interaction must be embedded into a more fundamental theory that will give rise to a $4 \times 4$ neutrino mass matrix in the bases $\nu_e,\nu_\mu,\nu_\tau,\nu_s$ diagonalized by a $4 \times 4$ unitarity matrix parametrized\footnote{We do not include in this counting the CP violating phases.} by three mixing angle between active-active states $\theta_{12}$, $\theta_{13}$, $\theta_{23}$ and three mixing angles between active-sterile states  $\theta_{1s}$, $\theta_{2s}$, $\theta_{3s}$. For simplicity we assume here that $\theta_{is}\ll 1$, so small as to neglect its effects. Even though this is a simplification which restricts the space of parameters we explore, it allows us to disentangle the effects due to the interaction from the effects due to the active-sterile mixing.

The couplings $\lambda_\alpha$ are free parameters of the model, which means that we have  an ample freedom of choice for our model. The most natural possibility is $\lambda_e=\lambda_\mu=\lambda_\tau$, since it preserves lepton universality.  However, also the case in which only $\lambda_\tau\ne0$ is very interesting:  even though it is not motivated by symmetry properties, we will see that it is only very weakly constrained by mesons decay. It can therefore lead to larger effects on the astrophysical fluxes without being excluded by present experiments. In our investigation  we therefore consider these two benchmark cases.

As mentioned in \cite{Fiorillo:2020jvy}, the cross section for the collision of sterile-active neutrinos exhibits a resonance in the t-channel. In fact, if a sterile neutrino with momentum $p$ collides with a fixed active neutrino, the former can decay producing an active neutrino and a scalar mediator which is then exchanged with the fixed active neutrino. The resonance condition $t=M_\varphi^2$ gives the following expressions for the energy of the final sterile and  of active neutrino:
\begin{equation} \label{eqdue}
E^i_s=\frac{m_i^2+m_s^2-M_\varphi^2}{2m_i}, E^i_a=\sqrt{p^2+m_s^2}+\frac{m_i^2-m_s^2+M_\varphi^2}{2m_i}
\end{equation}
where $m_i$ is the mass of the $i-$th active neutrino, $M_\varphi$ is the mass of the pseudoscalar $\varphi$
and $m_s$ is the mass of the sterile neutrino. 
Since $m_i\ll m_s, M_\varphi$,   from eq. \eqref{eqdue} it follows  that the resonance condition can be satisfied for positive energies if $m_s>M_\varphi$. 
If this condition is satisfied, the decay channel $\varphi\to \nu_s \nu$ is also kinematically suppressed. The amplitude for the process therefore depends critically on details of the model we have left unspecified. In fact, if the scalar mediator were completely stable, with no other decay channels, the t-resonance comes unregulated, giving rise to a non integrable pole in the differential cross section and a diverging total cross section. This situation is analogous to the divergence of the total cross section for Rutherford scattering. The regularization of this divergence in the case of a stable mediator particle depends, just as in the case of Rutherford scattering, on the transverse structure of the beam: if the radius of the beam is $a$, the cross section cannot exceed $\pi a^2$, so that the total cross section will saturate to this value.
It is however uncommon for a particle to be completely stable, if this stability does not descend from some specific property or conservation law. Therefore, It is unlikely that our mediator should be completely stable and  it may  have other decay channels, giving rise to a finite total decay rate $\Gamma$. This decay rate regularizes the divergence. Obviously,  this  implies a dependence on a new parameter $\Gamma$ into our work for the region $m_s>M_\varphi$. In what follows, we adopt the choice that the dominant decay channel is the decay into two active neutrinos \textit{via} a very small mixing angle. We will discuss in Section \ref{parasecret} the dependence of our results on this assumption.

\section{Constraints} \label{sec:constraints}

In the simple extension of the Standard Model under consideration, we introduce two new species of matter: the scalar field $\varphi$ and the sterile neutrino $\nu_s$. 
Our model is then  parametrized by the set 
\begin{equation}\label{setpar}
(\lambda_\alpha,\,M_\varphi,\,m_s)\,.
\end{equation}

Since this model is in principle subject to a number of constraints from  laboratory experiments, cosmology and  astrophysics, leading to a  restrictions 
of the free parameter space (\ref{setpar}), we take into account  the different constraints and we discuss them  in the following. As we will see, the results of this analysis suggest a region of interest in the parameters $10$ MeV$<m_s,M_\varphi<1$ GeV.

\subsection{Laboratory bounds}

\begin{figure} 
\includegraphics[width=0.45\textwidth]{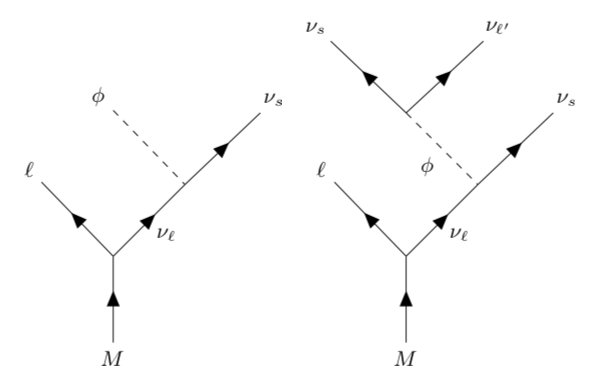}
\caption{Feynman diagrams for the new meson decay channels introduced by the interaction: time runs along the vertical axis.} \label{feyndiagrams}
\end{figure}

   It is well known that mesons can decay leptonically as $M\to \nu_\ell \ell $, where $M$ represents a meson ($\pi^+,\,K^+,\, D^+$) and $\ell=e,\mu,\tau$ depending on the meson. The interaction given in eq.\,(\ref{coupling}) opens the possibility of  new leptonic decay channels: $M\to \nu_s \ell \varphi$ and $M\to \nu_s \ell \overline{\nu}_{\ell'} \nu_s$. The Feynman diagrams for these new decay channels are shown in Figure \ref{feyndiagrams}. In this regard, we would like to remark that our assumption of small active-sterile mixing angle is crucial, since it prevents the appearance of new decay channels such as $M\to \nu_{\ell'} \ell \varphi$.

Concerning the process $M\to \nu_s \ell \varphi$,  it becomes possible only if the corresponding  
$\lambda_\ell \ne 0$; moreover, it is kinematically allowed only if 
\begin{equation}\label{kin1}
m_s+M_\varphi \lesssim m_M-m_\ell\,,
\end{equation}
where $m_M$ is the mass of the decaying meson and $m_\ell$ the mass of charged lepton $\ell$. A lower limit on $m_s+M_\varphi$ arises from Big Bang Nucleosynthesis as discussed below. 
Using the relation (\ref{kin1}),  in table\,(\ref{tab1}) we provide  the maximal allowed values for $m_s+M_\varphi$, namely $m_M-m_\ell$. 
\begin{table}[h!]
\begin{center}
\begin{tabular}{|r|c|c|}
\hline
Meson  & $(m_s+M_\varphi)_{\rm max}({\rm MeV})$  \\
\hline
$\pi^+\to e \varphi\nu_s $ & 140  \\
$\to \mu \varphi\nu_s $ & 35\\
$\to \tau \varphi\nu_s $ & --  \\
\hline
$K^+\to e \varphi\nu_s $ & 493  \\
$\to \mu \varphi\nu_s $ & 388  \\
$\to \tau \varphi\nu_s $ & --  \\
\hline
$D^+\to e \varphi\nu_s $ & 1870  \\
$\to \mu \varphi\nu_s $ & 1765  \\
$\to \tau \varphi\nu_s $ & 93  \\
\hline
\end{tabular}
\end{center}
\caption{New decay channels for light mesons induced by the interaction and relative maximal  allowed values for $m_s+M_\varphi$. When the numerical value is missing it means that the corresponding  process is kinematically forbidden.}
\label{tab1}
\end{table}
We observe that $\pi^+,K^+,D^+ \to e \varphi\nu_s $ and $K^+,D^+ \to \mu \varphi\nu_s $ have a large phase space available, while 
$\pi^+ \to \mu \varphi\nu_s $ and $D^+ \to \tau \varphi\nu_s $ are only marginally allowed.  At last,  the processes $\pi^+,K^+\to \tau \varphi\nu_s $ are not kinematically allowed.\\ 
In the active sector, experimental bounds on meson decay provide limits on  $\lambda_\ell$, see for example  \cite{Berryman:2018ogk}, where similar processes involving active neutrinos $\nu_\ell$  have been studied in details.
The main difference between our case and the one studied in \cite{Berryman:2018ogk} is that the relation (\ref{kin1}) is replaced with $M_\varphi \lesssim m_M-m_\ell$.  
In \cite{Berryman:2018ogk}, the limit on $\lambda_\mu$ from $K^+\to \mu \varphi\nu_s$ 
has been found to be  stronger with respect to the limit on $\lambda_e$ from $K^+\to e \varphi\nu_s$.  This depends on the more accurate experimental data available for the former process compared to the latter. Moreover, for masses of the scalar field $\varphi$ smaller than about $300$ MeV, the limits on $\lambda_{e,\mu}$ from $K^+$ decay are stronger compared to the one from $\pi^+$ and $D^+$ mesons decay.
Concerning  the $\lambda_\tau$ coupling, $\pi^+$ and $K^+$ can not provide information because of kinematics, see table\,(\ref{tab1}). We have explicitly analyzed the rate of process  $D^+ \to \tau \varphi\nu_s$, finding that it is very small compared with the experimental bounds. Therefore,  we find that for masses $M_\varphi$ and $m_s$ consistent with the cosmological constraints, $\lambda_\tau=1$ is always allowed.

In addition to three-body decay $M\to \nu_s \ell \varphi$ discussed above, there is a further  decay channel which occurs as a result of the new interaction: the four-body decay, $M\to \nu_s \ell \overline{\nu}_{\ell'} \nu_s$.  In this case the kinematics only constrains the mass of the sterile neutrino. In particular, the decay is kinematically allowed if
\begin{equation}\label{pippo}
2m_s \lesssim m_M-m_\ell\,.
\end{equation}
If $\lambda<1,$the rate for four-body decay will be smaller by a factor of $\lambda^2$ compared to the three-body decay\footnote{By $\lambda$ we mean the larger of the three couplings $\lambda_\alpha$.}. We therefore expect that the four-body decay can relevantly change the exclusions in the parameter space only if $\lambda=1$. Later we will explicitly verify the validity of the assertion.

We now discuss the explicit form of the decay rate of mesons through the new interactions. As a benchmark case we will discuss the Kaon decay into the muon channel: however, it is straightforward to obtain the decay rate for a different meson decay $M$ into the leptonic channel $\alpha$ by simply replacing in all the subsequent formulas $M_K$ by $m_M$ and $\mu$ by $\alpha$. 

We begin by examining the three body decay $K^+\to\mu\nu_s\varphi$. In the limit of vanishing active neutrino masses, the decay rate is
\begin{eqnarray}
&&d\Gamma_{K\to\mu\nu_s\varphi}=\\
&&\,\, =\frac{f_K^2 G_F^2 |\lambda_\mu|^2}{16 M_K (2\pi)^3}\int dE_p dE_k \frac{Q}{(M_K^2 + m_\mu^2 -2M_K E_p)^2},\nonumber	
\end{eqnarray}
where
\begin{eqnarray}
Q&=&8M_\varphi^2 [ 2(M_K E_k -E_p E_k +\mathbf{p}\cdot\mathbf{k}) \times \nonumber \\
&& (2E_p (M_K -E_p)-M_K E_p +m_\mu^2)-  \\
&&(M_K^2+m_\mu^2-2M_K E_p)(E_p E_k + \mathbf{p}\cdot\mathbf{k})].\nonumber 
\end{eqnarray}

We have defined with $p$, $q$, $k$ and $P$ respectively the $\mu$, $\varphi$, $\nu_s$ and $K$ four-momenta, and in bold face  their spatial three-momenta. $f_K$ is  the Kaon decay form factor.

With an analogous notations for the four body decay $K^+\to\mu\nu_s\nu_s\overline{\nu}_\ell'$, the decay rate can then be written: 

\begin{eqnarray}
&&d\Gamma_{K\to\mu\nu_s\nu_a\nu_s}=\frac{G_F^2 f_K^2 |\lambda_\mu|^2 (\sum_\alpha |\lambda_\alpha|^2)}{(2\pi)^6 M} \\
&&\qquad  \int\frac{|\mathbf{p}|^2 d|\mathbf{p}||\mathbf{q}|^2d|\mathbf{q}||\mathbf{k}|d|\mathbf{k}|d\cos\theta d\phi}{E_p E_q E_k |\mathbf{p}+\mathbf{q}|} \frac{q\cdot l Q'}{s^4\left[(q+l)^2 - M_\varphi^2\right]^2},\nonumber
\end{eqnarray}

 where $p$, $q$, $l$, $k$ and $P$  denote the four-momenta respectively of $\mu$, $\nu_s$, $\nu_\ell'$, $\nu_s$ and $K$. For convenience, we also defined $s=P-p$. We have also defined
 \begin{equation}
 Q'=4 k\cdot s p\cdot P s\cdot P -2 k\cdot s p\cdot s M_K^2 - 2 k\cdot P p\cdot P s^2+k\cdot p M_K^2 s^2.
 \end{equation}
  In this case, there are $5$ independent variables to parameterize the decay, which we choosed to be $|\mathbf{p}|$, $|\mathbf{q}|$, $|\mathbf{k}|$, the angle $\theta$ between $\mathbf{p}$ and $\mathbf{q}$, and the azimuthal angle $\phi$ between $\mathbf{k}$ and the plane determined by $\mathbf{p}$ and $\mathbf{q}$. In case $\mathbf{p}$ and $\mathbf{q}$ were collinear, this should be interpreted as the azimuthal angle around the direction of $\mathbf{p}$

Both the processes $K^+\to \mu \varphi\nu_s $ and $K^+\to\mu\nu_s\nu_s\overline{\nu}_\ell'$ should be observed as $K\to\mu+\mbox{missing energy}$. The closer Kaon decay process that is reported into PDG \cite{PhysRevD.98.030001} \ is $K\to\mu\nu\overline{\nu}\nu$ that can be used to constrain our processes. Therefore, we impose that  the branching ratio to this channels should be smaller than $2.4\times10^{-6}$ \cite{PhysRevD.98.030001}.

\begin{figure}[h!]
\centering
\includegraphics[width=0.45\textwidth]{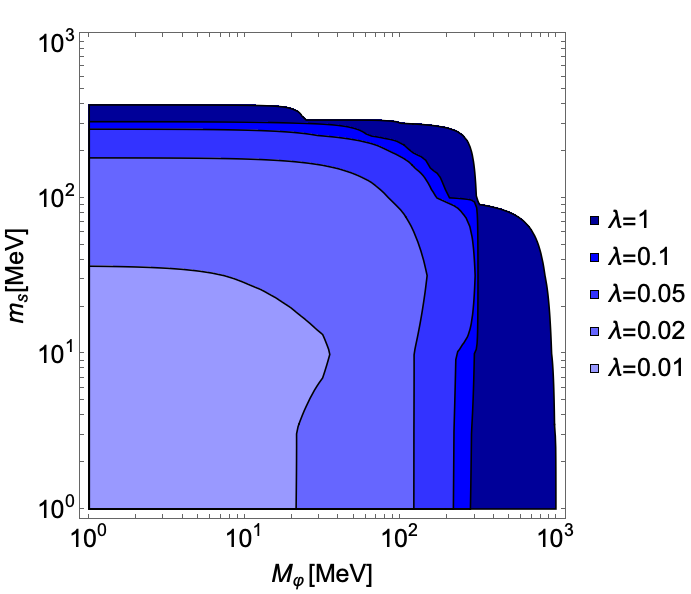}
\caption{Exclusion contours in the $M_\varphi-m_s$ plane for different values of the coupling  $\lambda=\lambda_e=\lambda_\mu=\lambda_\tau$, for the choice of equal flavor coupling: the region below the contours is excluded.}\label{fig1}
\end{figure}

As mentioned in section \ref{sec:model}, a reasonable choice for a  qualitative picture of the general case is to take $\lambda_e=\lambda_\mu=\lambda_\tau=\lambda$. In Figure \ref{fig1} we consider this case and we show the region excluded by Kaon decay in the $M_\varphi - m_s$ plane for various values of the coupling.
From figure (\ref{fig1}) we observe that if 
\begin{equation}\label{lowerlimitmass}
\lambda_{} \ge 0.01\, \, \text{and} \, (m_s \,\mbox{or}\, M_\varphi) \gtrsim 30\,MeV \,, 
\end{equation}
then the correction to Kaon decay is within the experimental bound.

The four-body decay channel only produces a bump in the right part of the exclusion contours, corresponding to a roughly horizontal line of exclusion that only constrains $m_s$, as expected from our previous considerations: just as we had deduced, this bump is only relevant for $\lambda=1$.

The only case in which the results are drastically different from the choice of equal couplings for the three flavors is  the one in which $\lambda_\tau\ne 0$ and the other two couplings are much smaller than it. In fact,  as we mentioned above, this case is unconstrained from meson physics and  even for value of $\lambda_\tau\sim \mathcal{O}(1)$ 
the only relevant bound in the $M_\varphi - m_s$ plane comes from Big Bang Nucleosynthesis as discussed below.

\subsection{Cosmological bounds}
In addition to laboratory bounds, there could be additional constraints coming  from cosmology at different epochs of the Universe. 
A first constraint comes from the Big Bang Nucleosynthesis (BBN) epoch  and specifically  from the number of relativistic degrees of freedom. In particular, the requirement is that there are no extra relativistic species (apart from the ones predicted from the Standard Model) at the moment of the BBN. This naturally happens if the new introduced  species $\varphi$ and $\nu_s$ are non relativistic and in thermal equilibrium before and during BBN. Indeed, their distributions will be then Boltzmann suppressed by a factor of $e^{-m/T_{\rm BBN}}$, where $T_{\rm BBN}\simeq 1$ MeV, and will not count as extra degree of freedom.
Another  constraint comes from the requirement that the new interaction does not affect the free-streaming nature (non interacting) of the active neutrinos  at the time of the formation of the Cosmic Microwave Background (CMB). 

A full discussion of the cosmological bounds would require explicit solution of the evolution equations for all the relevant species and is outside the scope of this work. We will limit ourselves to an order of magnitude estimate of the rates of the relevant processes to get a clear idea of which portion of the parameter space is effectively constrained. Therefore, we did not distinguish between the three different coupling $\lambda_\alpha$  but  we used an effective coupling $\lambda$. This latter have been  chosen as the largest coupling between the three and so the most relevant.

\subsection*{ Big Bang Nucleosynthesis}
As mentioned above, the requirements to not affect the BBN yields are that the newly introduced species are non relativistic at the time of the BBN and that they remain in kinetic and chemical equilibrium throughout the passage from relativistic to non relativistic. The first requirement is naturally met if both $M_\varphi$ and $m_s$ are chosen to be larger than about $10$ MeV: in this way, the Boltzmann factor is smaller than $10^{-4}$ and we can safely assume that the species are non relativistic. Concerning the second requirement, is necessary to compare the rates of the processes responsible for  the equilibrium with the rate of the expansion of the Universe, in order to determine the temperature of decoupling at which such processes become irrelevant.
\begin{itemize}
\item $\nu_\alpha \nu_s \to \nu_\alpha \nu_s$ and $\nu_s \nu_s \to \nu_\alpha \nu _\alpha$: the cross section of these processes, mediated by $\varphi$, can be estimated as $\frac{\lambda^4 T^2}{M_\varphi^4}$
. If we assume a non relativistic distribution for the particles involved, consistently with our requirement that the newly introduced species decouples while being non relativistic, the sterile number density grows as $n\sim (Tm_s)^{3/2} e^{-m_s/T}$. The decoupling temperature will be set by the condition $n\sigma\sim H$, where $H$ is the Hubble parameter, which translates into the condition $n \sigma\sim \frac{T^2}{M_{Pl}}$, where $M_{Pl}$ is the Planck mass. This leads to the relation $$\left(\frac{T}{m_s}\right)^{3/2}e^{-m_s/T}\sim\frac{M_\varphi^4}{M_{Pl}m_s^3 \lambda^4}.$$ An approximate solution for this equation, in the regime in which $M_{Pl} m_s^3 \lambda^4 \gg M_\varphi^4$, is then $$T_{s}\sim\frac{m_s}{\log\left[\frac{M_{Pl} m_s^3 \lambda^4}{M_\varphi^4}\right]}.$$The factor in the denominator depends only logarithmically, and therefore very weakly, on the parameters, and for typical values of the masses between $10$ MeV and $1$ GeV and $\lambda$ between $0.01$ and $1$ is of the order of $10$ to $100$. Therefore we find that the decoupling temperature is of the order of $\frac{m_s}{10}$, which means that these processes are able to maintain both kinetic and chemical equilibrium between active and sterile neutrinos even after the latter have become non relativistic, ensuring the Boltzmann suppression of this species.

\item $\varphi \varphi \to \nu_s \nu_s$ or $\varphi \varphi \to \nu_\alpha \nu_\alpha$: the most efficient is the first  process mediated by active neutrino. The cross section is estimated as $\frac{\lambda^4}{m_\alpha^2}$, where $m_\alpha$ is the active neutrino mass, and if we again assume a non relativistic distribution for the scalar particles we find the condition $$\left(\frac{T}{M_\varphi}\right)^{-1/2} e^{-M_\varphi/T}\sim\frac{m_\alpha^2}{M_{Pl} M_\varphi \lambda^4}.$$ An approximate expression for the decoupling temperature, in the regime $m_\alpha^2 \ll M_\varphi M_{Pl} \lambda^4$, is $$T_\varphi \sim \frac{M_\varphi}{\log\left[\frac{M_\varphi M_{Pl} \lambda^4}{m_\alpha^2}\right]}.$$ Since the factor in the denominator is typically of order $10$ to $100$, we find again that the scalar particles remain in equilibrium throughout their passage from relativistic to non relativistic, and they therefore become Boltzmann suppressed, not counting as radiative degrees of freedom.
\end{itemize}

In summary, for the parameter space which is of interest to us, both scalar and sterile particles remain in kinetic and chemical equilibrium throughout the primordial nucleosynthesis. By taking them sufficiently massive, namely 
\begin{equation}
m_s\gtrsim 10\,\mbox{MeV and }\,M_\varphi \gtrsim 10\,\mbox{MeV},
\end{equation}
we can safely satisfy BBN limits, since the newly introduced particles are so massive that they are Boltzmann suppressed during BBN.

\subsection*{Cosmic Microwave Background}

At the time of formation of the Cosmic Microwave Background (CMB), sterile neutrinos and scalar particles have long disappeared. Active neutrinos, though, can still secretly  interact through the reactions $\nu_\alpha \nu_{\alpha'} \to \nu_\beta \nu_{\beta'} $. 
There are in principle two ways in which this interaction can proceed through the new interaction: either the mixing angle between active and sterile neutrinos is sufficiently large, so that the process $\nu_\alpha \nu_s \to \nu_\alpha \nu_s$ can be converted \textit{via} mixing to a four active neutrino process; or the process can happen at next-to-leading order \textit{via} the box diagram.
We will only analyze the latter process, since we have assumed very small mixing angles between active and sterile neutrinos. We assume a relativistic distribution for the active neutrinos. The cross section can be estimated in order of magnitude as $\frac{\lambda^8 T^{10}}{M_\varphi^8 m_s^4}$, so that the decoupling temperature for this process is
\begin{align*}
T^{\rm dec}_{\nu_\alpha \nu_\alpha'} &=\left(  \frac{M_\varphi^8 m_s^4}{\lambda^8 M_{\rm pl}}  \right)^{1/11}\simeq \\
&\simeq 10^5 \text{eV} \left(\frac{M_\varphi}{10 \text{MeV}}\right)^{8/11}\left(\frac{m_s}{10 \text{MeV}}\right)^{4/11}\lambda^{-8/11}.
\end{align*}
In order to guarantee free-streaming active neutrinos at CMB time, $T^{\rm dec}$ has to be larger  than the temperature of  CMB formation, around  $1$ eV. We have checked that  this is the case for all the parameter space we considered.

\subsection{Astrophysical bounds}

Another possible constraint we should take into account comes from the analysis of neutrino fluxes from supernovae. This kind of constraints have been recently taken into account in the analysis of active-active secret interactions \cite{Mastrototaro:2019vug,Shalgar:2019rqe}. In fact, since neutrinos in the supernovae core have energies of order of tens or hundreds of MeV, they are sufficiently energetic to produce non relativistic sterile neutrinos. If these sterile neutrinos interact sufficiently weakly with the active neutrinos in the core, they could escape the supernova giving rise to an observable energy loss. The conditions for this to happen are two: in the first place, the mean free path of the sterile neutrino inside the core, namely $\left(\sigma_{sa} n_a\right)^{-1}$, with $\sigma_{sa}$ the cross section and $n_a$ the number density of active neutrinos in the core, should be larger than the radius of the supernova core, typically around $10$ km. The cross section is evaluated for active neutrinos with typical energies of tenth of MeV and sterile neutrinos at rest. The number density $n_a$ can be estimated assuming a thermal distribution $f(E)$ of active neutrinos, with a typical temperature of tenth of MeV, as in \cite{Mastrototaro:2019vug}.

The second condition to be verified is that  sterile neutrinos should be copiously produced in the supernova core and that the energy injected into sterile neutrinos can  exceed the threshold luminosity $L_s \simeq 2\times 10^{52}$ erg/s (namely $L_s\simeq 8.2\times 10^{36}$ MeV$^2$ in natural units, which we have used throughout this work)  for the supernova SN 1987A \cite{Mastrototaro:2019vug}.
We then estimate the luminosity  in the proposed model as:
\begin{equation*}
L_s=\int \frac{d\sigma_{a\to s}}{dE} E dE f(E',r) f(E'',r) dE' dE'' 4\pi r^2 dr,
\end{equation*}
where $f(E,r)$ is the distribution of active neutrinos inside the core of the supernova. The temperature profile $T(r)$ is taken from \cite{Mastrototaro:2019vug}.

 The model under consideration could be in conflict  with SN 1987A data if both the above conditions would simultaneously met. We have numerically verified that this situation never occurs for all the parameter space we considered, with $m_s$ and $M$ larger than $10$ MeV. Indeed, for large values of the coupling $\lambda$ the energy injected into sterile sector can easily exceed the threshold indicated above, but the interaction between sterile and active neutrinos is so strong that the mean free path is much shorter than the supernova dimensions. For small values of $\lambda$ we encounter  the opposite situation  where, even if sterile neutrinos are practically free to escape the supernova, they are produced in too small amounts to be observable.  We can therefore deduce that the model we consider is not constrained by supernova data.

\section{Neutrino fluxes} \label{sec:fluxes}
Active-sterile neutrino interaction can become relevant at very different energy scales depending on the mass of the scalar mediator $\varphi$: roughly we expect the energy scale at which the process of absorption over neutrinos from the Cosmic Neutrino Background (CNB) happens resonantly \footnote{Of course if the sterile mass is too large it can kinematically forbid the process: in determining the energy at which the absorption is most relevant one should take this factor into account.} at energies around $M_\varphi^2 / m_\alpha$.  For an active neutrino mass of $0.1$ eV, we find that this energy scale can range from PeV to energies of order $10^4$ PeV  in the selected parameter  space. Close to the PeV scale the dominant source of neutrinos is expected to be constituted by galactic and extragalactic astrophysical sources, among which we mention Active Galactic Nuclei (AGN) and Gamma Ray Bursts (GRB). The details of the emitted neutrino spectra are sensitive to the physics of the sources. However, it is known that a good fit to the observed IceCube data in the region below the PeV is represented by a simple power law spectrum. Therefore, in this range of energy,  we limit our  discussion on the effect of the new interaction  on a power law spectrum with parameters obtained by the fit to the IceCube data given in \cite{Williams:2018kpe}.

At higher energies, from $100$ PeV, there are no experimental data on the neutrino flux. It is expected that a dominant source of neutrinos should have cosmogenic origin. On the other hand, recent studies have shown that a competing source of neutrinos could still be of astrophysical nature, provided for example by blazars \cite{Murase:2014foa} and Flat Spectrum Radio Quasars (FSRQ) \cite{Righi:2020ufi}. 

The new interaction under consideration produces however effects which are qualitatively the same on all these fluxes. For simplicity, we stick to the treatment adopted in our recent paper \cite{Fiorillo:2020jvy} considering the effects of the new interaction on cosmogenic neutrino fluxes.

\subsection{Without secret interaction}
\subsection*{Power Law}
We consider a collection of astrophysical neutrino sources, each one producing a power law spectrum per unit solid angle\footnote{If the source is anisotropic, the spectrum is evaluated in the direction of the Earth.} in energy
\begin{equation}
\frac{dN_\nu}{dEdtd\Omega}=g(E) = \mathcal{N} \, E^{-\gamma},
\end{equation}
where $g\equiv \phi_{\nu_e}+\phi_{\nu_\mu}+\phi_{\nu_\tau}+\phi_{\overline{\nu}_e}+\phi_{\overline{\nu}_\mu}+\phi_{\overline{\nu}_\tau}$ and $\gamma$ is the spectral index. The IceCube analysis gives as best fit value for the throughgoing muons data set $\gamma=2.28$ \cite{Schneider:2019ayi}. Due the similarity with the cosmogenic neutrino production, we found  convenient to adopt the Star Forming rate $\rho(z)$ \cite{Hopkins:2006bw} for the cosmological evolution of these sources, where $\rho(z)$ is defined as the comoving number density.
The normalization $\mathcal{N}$ is chosen in such a way as to reproduce the best fit for the diffuse neutrino flux measured by the IceCube Collaboration in the throughgoing muons data sample. The flux arriving at Earth from the point-like source, expressed in terms of $g(E)$, is
\begin{equation}
\frac{d\phi}{dE}=\frac{g\left[E(1+z)\right]}{r^2(z)},
\end{equation}
where $r(z)=\int_0^z \frac{dz'}{H(z')}$.\\
Therefore, the diffuse astrophysical spectrum is written as
\begin{equation}
\frac{d\phi_{\nu}}{dEd\Omega}= \int\frac{dz'}{H(z')} \rho(z') g[E(1+z')].
\end{equation}
We assume for definiteness a flavor structure at the source $(1:2:0)$, corresponding to pion beam sources.

Throughout our analysis we  use the best fit values from the NuFit 3.2 global fit data  for the  active oscillation parameters \cite{Esteban:2016qun}, assuming normal neutrino mass ordering.

\vskip5.mm
\subsection*{Cosmogenic}
Cosmogenic neutrinos are produced by the scattering of high energy protons from the cosmic rays with the CMB photons. The production of cosmogenic neutrinos is quantitatively studied, for example, in \cite{Ahlers:2012rz}. In our previous work \cite{Fiorillo:2020jvy} we showed that their results can be reproduced by parameterizing the cosmogenic neutrino spectrum as

\begin{equation}
\frac{d\phi_{\nu}}{dEd\Omega}= \int\frac{dz'}{H(z')} \rho(z') f[E(1+z')],
\end{equation}
where $\rho(z)$ is the Star Forming rate \cite{Hopkins:2006bw}. We refer the reader to \cite{Fiorillo:2020jvy} for the method of determination of the function $f(E)$ describing the energy spectrum. 

For cosmogenic neutrinos we again assume a flavor structure at the source $(1:2:0)$.

\vskip25.mm

\subsection{With secret interaction} \label{parasecret}
Because of the secret interactions, active neutrinos can collide with active neutrinos from the CNB, producing sterile neutrinos and thereby causing a depletion of the flux observable at Earth. The transport equation for active neutrinos is in principle coupled to the transport equation for sterile neutrinos, since the secret interactions produce sterile neutrinos which can in turn collide with other CNB neutrinos to regenerate part of the original flux. The form of these equations has been given in \cite{Fiorillo:2020jvy}, and we reproduce it here for the generalized multiflavor case. We define   with $\Phi_{i} (z,E)$ the flux of active neutrinos in the $i$th ($i=1,2,3$) mass eigenstate per unit energy interval per unit solid angle  at a redshift $z$, while  $\Phi_{s} (z,E)$  denotes the flux of sterile neutrinos, where in absence of mixing the sterile mass  eigenstate is indicated with $s$. The flux at Earth is connected with the flux at generic redshift by the relation $\frac{d\phi_{\nu}}{dEd\Omega}=\Phi(0,E)$. We write separate equations for the mass eigenstates because, as discussed in \cite{Fiorillo:2020jvy},   the propagation is diagonal in the mass eigenstates, given the path between collisions much larger than the oscillation lengths.  In other words, due to the very fast oscillations caused by mixing, in between two collisions, a neutrino decoheres to mass eigenstates.

The transport equations take the form:
\begin{eqnarray}
&&H(z)(1+z)\left(\frac{\partial \Phi_i (z,E)}{\partial z} + \frac{\partial \Phi_i(z,E)}{\partial E}\frac{E}{1+z}\right)= \nonumber\\ &&n(z) \sigma_i \Phi_i (z,E)\nonumber\\&&- \int dE' \Phi_s(z,E') \frac{d\sigma_{sa}}{dE}(E'\to E) n(z)\nonumber \\&&-\rho(z)(1+z) f(E) \xi_i,\label{Ta}
\end{eqnarray}
where $f(E)$ is the neutrino spectrum produced at the source and $\xi_i$ is the fraction of neutrinos produced at the source in the $i$th mass eigenstate. Similarly, for the sterile flux we write:
\begin{eqnarray}
&&H(z)(1+z)\left(\frac{\partial \Phi_s(z,E)}{\partial z} + \frac{\partial \Phi_s(z,E)}{\partial E}\frac{E}{1+z}\right)= \nonumber\\ &&n(z) \sigma_s \Phi_s (z,E) \nonumber\\ &&- \sum_i \int dE' \Phi_i(z,E') \frac{d\sigma_{as}}{dE}(E'\to E) n(z)\nonumber\\ &&- \int dE' \Phi_s(z,E') \frac{d\sigma_{ss}}{dE}(E'\to E) n(z).\label{Ts}
\end{eqnarray}
For convenience, we have denoted by $\sigma_i$ and $\sigma_s$ the cross sections for the collision of an $i$th mass eigenstate and a sterile neutrino, respectively, with a CNB neutrino. Correspondingly, $\frac{d\sigma_{\alpha\beta}}{dE}(E'\to E)$ is the cross section for the production of a $\beta$ neutrino with energy $E$ after the collision of a $\alpha$ neutrino with energy $E'$ with a CNB neutrino. However, it is important to notice that if $m_s>M_\varphi$, a further process needs to be taken into account corresponding to the possible decay of the sterile neutrinos into an active neutrino and a scalar mediator. We refer the reader to Appendix B for the mathematical treatment of this case.

 If the regeneration processes play an important role, the task of determining the effect of the interaction is computationally expensive, since it requires the numerical solution of the system of four coupled partial integro-differential equations  

In our previous paper \cite{Fiorillo:2020jvy},  we found that the regeneration was unimportant  for a limited region of the parameter space, with masses of sterile neutrino and scalar mediator around $250$ MeV. Here we have analyzed this question more thoroughly, taking in consideration a wider parameter space. We have adopted a perturbative approach in which the regeneration processes are treated as a perturbation and  we have tested its validity \textit{a posteriori} by comparing the perturbation induced by regeneration with the unperturbed flux. 

We find that both cosmogenic and astrophysical fluxes are practically unaffected by regeneration. The physical reason behind this behavior  is connected with the cosmological evolution of the sources, and in particular with the fact that the sources are distributed at various redshifts. In fact, while neutrinos produced at high redshifts, with $z\gg 0.1$, are severely suppressed due to the absorption on the CNB, neutrinos produced at low redshifts are only weakly absorbed. Thus the flux has always a component, produced at low redshift, which is roughly unabsorbed and which dominates against the small regenerated flux produced at high redshifts. The perturbative approach shows in fact that the corrections coming from regeneration, both for cosmogenic and astrophysical fluxes, are typically not larger than about $10\%$. This conclusion is not reached in the case $m_s>M_\varphi$, where sterile neutrino decays are important, as described in Appendix B. In this case, we find that the results of the first order perturbation theory may cause small but non negligible changes to the spectrum. For this reason, in the following the regime $m_s>M_\varphi$ has been treated taking into account regeneration perturbatively to first order.

\begin{figure}[h!]
\centering
\includegraphics[width=0.45\textwidth]{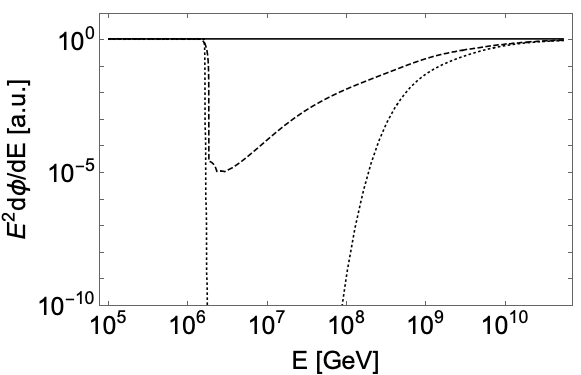}\\
\includegraphics[width=0.45\textwidth]{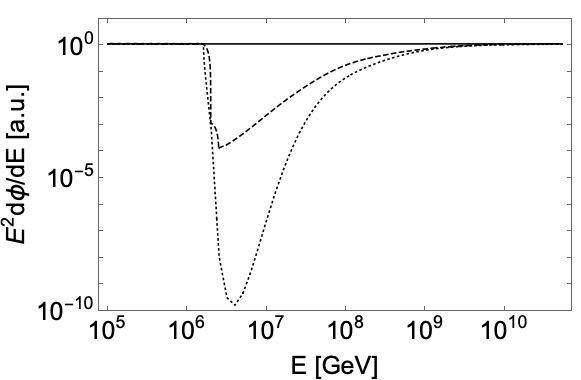}
\caption{Comparison between the spectra with pure absorption (dotted lines) and with both absorption and regeneration (dashed lines) for an $E^{-2}$ flux produced by a source at redshift $0.1$ (top panel) and $0.01$ (bottom panel). The thick line is the unabsorbed spectrum, reproduced for reference. The sterile and scalar mediator masses are fixed to the benchmark values of $10$ MeV; the coupling $\lambda$ is chosen as $1$ for the tau neutrinos.}\label{fixedred}
\end{figure}

The negligible effect of regeneration is therefore connected with the presence of sources at small redshifts, masquerading the regenerated flux. Thus we expect that, for point-like sources localized at large redshifts, regeneration effects should instead be non negligible. Even though  IceCube has identified so far a single realistic candidate of point-like astrophysical source, in the future one may expect noticeable improvements in this respect. Therefore, it might be interesting to have a qualitative idea of the effect of regeneration on the neutrino spectra from point-like sources. In Figure \ref{fixedred} we show the spectra expected at Earth for a generic source at  two fixed redshift values $z$, namely  $0.1$  and $0.01$,  with an $E^{-2}$ reference spectrum. The effects of regeneration are, as expected, more important for larger redshifts of the source and can drastically change the results. 

%A final comment on the regeneration effects is necessary in relation to the  very small active-sterile mixing angle we adopted. In fact, as we mentioned above, in the regime $m_s>M_\varphi$, the mediator decay into an active and sterile neutrino is inhibited. Under these conditions we consider a mediator decay into two active neutrinos \textit{via} a very small active-sterile mixing angle. We evaluated  the dependence of  negligible regeneration on the value of the mixing angle. In general large active-sterile mixing angles would lead to smaller regeneration effects, because they increase the decay rate of the mediator. On the other hand we find that even very small mixing angles still produce no relevant regeneration effects. Therefore, our conclusion that regeneration can be neglected for diffuse fluxes is still valid. Since the active-active cross section does not exhibit any t-channel resonance, it does not depend on the precise value of the mixing angle, making our results roughly independent of this assumption.

In the following, since we  only deal with cosmogenic and astrophysical neutrino fluxes, we  neglect the regeneration processes, so that the transport equation for the sole active neutrinos is given by
\begin{eqnarray}
&&H(z) (1+z)\left[\frac{\partial \Phi_{i}}{\partial z} +\frac{\partial \Phi_{i}}{\partial E} \frac{E}{1+z}\right] = n(z) \sigma_{i}(E) \Phi_{i}(E) + \nonumber \\ 
&&-\rho(z) f(E)\xi_i\label{te}
\end{eqnarray}
In this equation $\sigma_{i}$ is the cross section of the process $\nu_i +\nu_j\to \nu_s+\nu_s$, namely
\begin{eqnarray}
\sigma_{i}=\frac{1}{64\pi I^2} \sum_j \int_{t_1}^{t_2} |\mathcal{M}_{i j\to ss}|^2 (s,t) dt \label{eqsaass}
\end{eqnarray}
in terms of the Mandelstam invariants $s=(p+l)^2$, $t=(p-k)^2$ and $u=(p-q)^2$ with $p$, $l$, $k$ and $q$ the momentum of the two active neutrinos and the two sterile neutrinos, respectively. Moreover 

\begin{equation}
t_{1,2}=m^2+m_s^2-\frac{s}{2}\pm\sqrt{s}\sqrt{\frac{s}{4}-m_s^2}\, ,
\end{equation}
and
\begin{equation}
I=\sqrt{\frac{2m^4+s^2-4sm}{2}}
\end{equation}
where $m$ is the mass of the active neutrino $\nu$ of CNB.
The squared amplitude is given by
\begin{eqnarray} \label{amplit}
&&|\mathcal{M}_{ij\to ss}|^2 =|\sum_{\alpha,\beta} U_{\alpha i}^* U_{\beta j}^* \lambda_\alpha \lambda_\beta|^2 \nonumber \\
&& \times \left[ \frac{[t-(m-m_s)^2]^2}{(t-M_\varphi^2)^2+\Gamma^2 M_\varphi^2}
+ \frac{[u-(m-ms)^2]^2}{(u-M_\varphi^2)^2+\Gamma^2 M_\varphi^2} \right.\nonumber\\
&&- \frac{2[(t-M_\varphi^2)(u-M_\varphi^2)+\Gamma^2 M_\varphi^2]}{[(t-M_\varphi^2)^2+\Gamma^2 M_\varphi^2] [(u-M_\varphi^2)^2+\Gamma^2 M_\varphi^2]}  \nonumber\\
&&\times \left(\frac{(t-m^2-m_s^2)^2}{4} +\frac{(u-m^2-m_s^2)^2}{4}  \right.\nonumber\\
&&-\left. \left. \frac{s^2}{4}+s(m^2+m_s^2-m \, m_s) -2m^2 m_s^2 \right)  \right]
\end{eqnarray}
where $\Gamma$ is the decay rate of the scalar mediator and $M_\varphi$ is its mass.

Therefore eq.(\ref{te})  contains only an absorption term and,   for the astrophysical power law neutrino flux, it admits  an analytical solution for the flux at Earth given by
\begin{eqnarray} \label{abs}
\Phi_{{i}} (E)=\int_0^{+\infty} \frac{dz}{H(z)} \rho(z) g\left[ E(1+z)\right] \times \nonumber \\  \exp\left[-\int_0^z \frac{dz'}{H(z') (1+z')} n(z') \sigma_{\nu_i}\left[E(1+z')\right]\right] \xi_i\,.
\end{eqnarray}
 For cosmogenic neutrino fluxes the solution is identical with we consider the function $f(E)$ in place of $g(E)$.

%%%%%%%%%%%%%%%%%%%%%%%%%%%%%%%%%%%%%%%%%%%%

\section{Results} \label{sec:results}

\subsection{Power law}

We start our discussion of the results with the case of a power law astrophysical spectrum in the energy range roughly below $100$ PeV. In this region the effects of active-sterile interaction can be detected only if the sterile mass is sufficiently low that the process is not kinematically forbidden. As in the previous sections, we distinguish between the two possibilities: either $\lambda_e=\lambda_\mu=\lambda_\tau=\lambda_{af}$ (where $af$ denotes all flavors), or $\lambda_e=\lambda_\mu=0$ and $\lambda_\tau \neq 0$. 

In the first case, the Kaon decay strongly constrains the possible values of the coupling. In particular, we find that the optimal choice to have noticeable effects below $100$ PeV is to have small sterile masses, large scalar masses and $\lambda_{af}=1$. We take as benchmark values $m_s=10$ MeV and $M_\varphi = 1$ GeV.  

\begin{figure}[h!]
\centering
\includegraphics[width=0.5\textwidth]{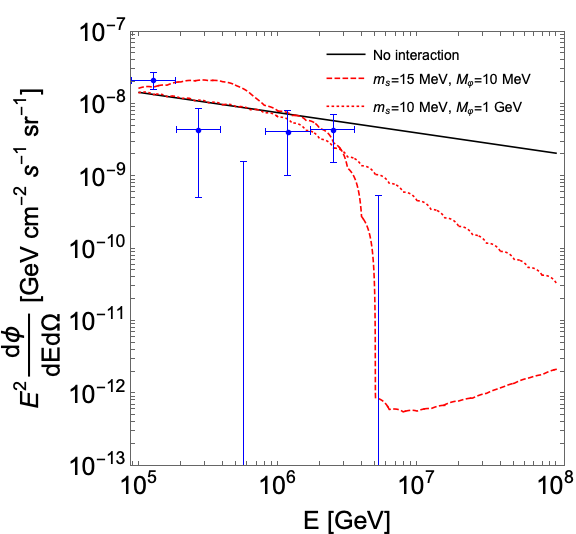}
\caption{Astrophysical all flavor neutrino power law spectra: the thick line is the flux with no interaction, while the dotted line corresponds to the case of $\lambda_{af}=1$ and the dashed one denotes the case  $\lambda_\tau=1$, as described in the text. The experimental points are the IceCube HESE data.}\label{plIC}
\end{figure}

In the second case, in which the mediator only couples to tau neutrinos, the constraints from meson decays are irrelevant and we can also consider lower masses for $M_\varphi$.  In order to maximize the effect in this energy range,  we  have chosen the benchmark values of $m_s=15$ MeV, $M_\varphi=10$ MeV and $\lambda_\tau=1$, as represented in
 Fig, \ref{plIC} where we show the neutrino spectra after the new interaction for both the choices of $\lambda$, together with the IceCube HESE data \cite{Peretti:2019vsj}.  From this Figure we can infer that the second possibility is already testable using IceCube data while the first  case is essentially undistinguishable from the power-law in the energy range probed by IceCube.

An interesting aspect revealed by Figure \ref{plIC} is that the new interaction causes a cutoff-like feature in the spectrum in the range between $1$ PeV and $10$ PeV. In fact, the second case with only $\lambda_\tau$,  shows a sudden drop of the flux at the energy at which the process $\nu_\alpha \nu_\beta \to \nu_s \nu_s$ becomes kinematically allowed.

We also emphasize that, in the case $m_s=15$ MeV, $M_\varphi=10$ MeV, where regeneration may slightly change the results because of the presence of sterile neutrino decays, in addition to the absorption, there is a small pile up of neutrinos in the region between $100$ TeV and $1$ PeV. As mentioned before, this result has been obtained using first order perturbation theory for the treatment of the regeneration term.

\begin{figure}[h!]
\centering
\includegraphics[width=0.5\textwidth]{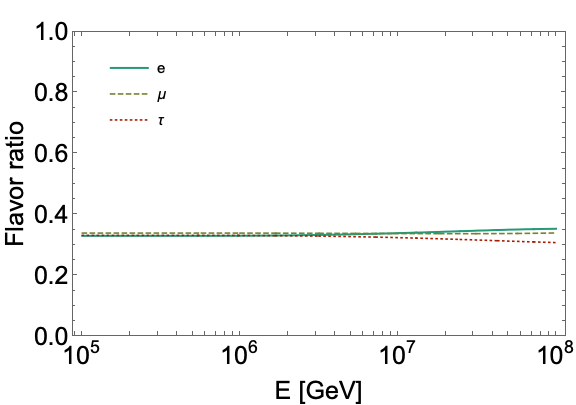}\\
\includegraphics[width=0.5\textwidth]{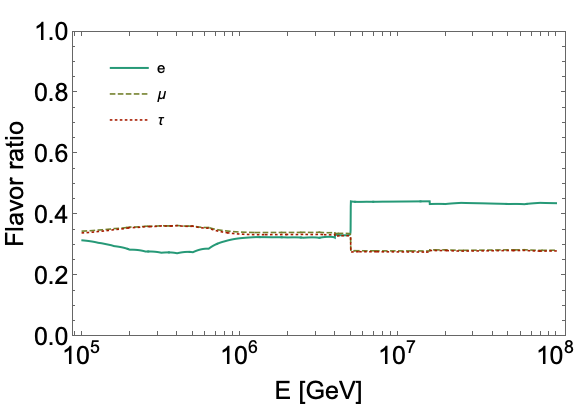}
\caption{Flavor ratio at Earth as a function of the energy for the first benchmark case in the text ($m_s=10$ MeV, $M_\varphi=300$ MeV, $\lambda_e=\lambda_\mu=\lambda_\tau=1$) (top pannel) and 
second benchmark case in the text ($m_s=15$ MeV, $M_\varphi=10$ MeV, $\lambda_\tau=1$) (bottom pannel).}\label{frIC2}
\end{figure}

The effects of the new interaction can also cause significant changes in the flavor structure of the spectrum since  the induced depletion acts differently on each flavor modifying  the flavor ratio, namely the fraction of electron, muon and tau neutrino fluxes. Since the depletion is energy dependent, the result will be an energy dependent flavor ratio. In Figures \ref{frIC2} we show the flavor ratios as a function of the energy for the two  cases,  $\lambda_{af}$  and $\lambda_\tau$ respectively. We see that the case of only $\lambda_\tau$  has a threshold behavior with a sudden change of the flavor ratio. This change is quite relevant, especially when compared with the change in the  case $\lambda_{af}$. We remind the reader that we assumed a flavor ratio at the source $(1:2:0)$: at low energies, where the effects of the interaction are inactive, we recover the typical flavor structure $(1:1:1)$ at the Earth as expected.

\begin{figure}[h!]
\centering
\includegraphics[width=0.35\textwidth]{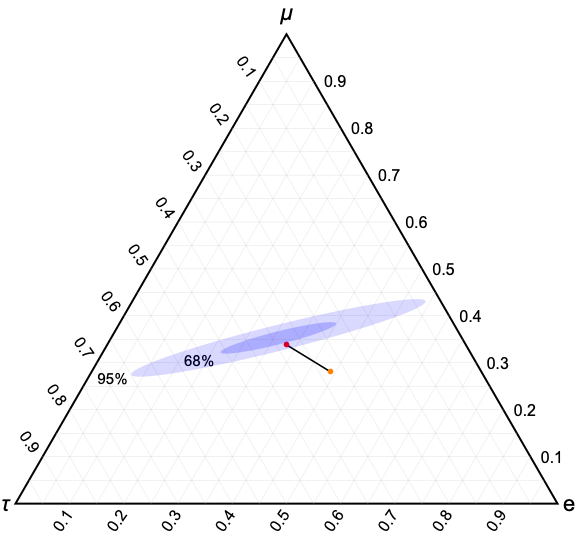}
\caption{Flavor ratio, reproduced in the flavor triangle, for varying energy for the second case in the text ($m_s=15$ MeV, $M_\varphi=10$ MeV, $\lambda_\tau=1$). The red and orange points correspond to an energy of $10^5$ GeV and $10^8$ GeV, respectively. The forecasted sensitivity of IceCube-Gen2 is shown as well.}\label{trIC2}
\end{figure}

The effects of secret interaction on  the flavor structure of the spectrum, namely the modifications of the flavor ratio,  can also be represented in the \textit{flavor triangle}: we show this for case $\lambda_\tau=1$, which has the largest effect, in Figure \ref{trIC2}.
\\The red and the orange points correspond to an energy of $10^5$ GeV and $10^8$ GeV, respectively. The flavor sensitivity which has been forecasted for IceCube-Gen2 \cite{Bustamante:2019mar}, in the case of pion beam sources with a flavor ratio $(1:2:0)$ at the source, has been shown as well. The triangle representation suggests the possibility  that future experiments might be able to unveil a different flavor structure  possibly caused by active-sterile secret interactions. It is worth noticing that this change induced by the interaction is also dominant with respect to the uncertainty due to the mixing parameters.
\\A fundamental feature of the change in flavor induced by secret interactions is that it has a unique energy dependence, which descends from the resonances and thresholds of the interaction. Since the data from IceCube-Gen2 might allow in the future to investigate the interplay between flavor and energy, this is a result which might be of experimental interest.

\subsection{Cosmogenic}

\begin{figure}[h!]
\centering
\includegraphics[width=0.5\textwidth]{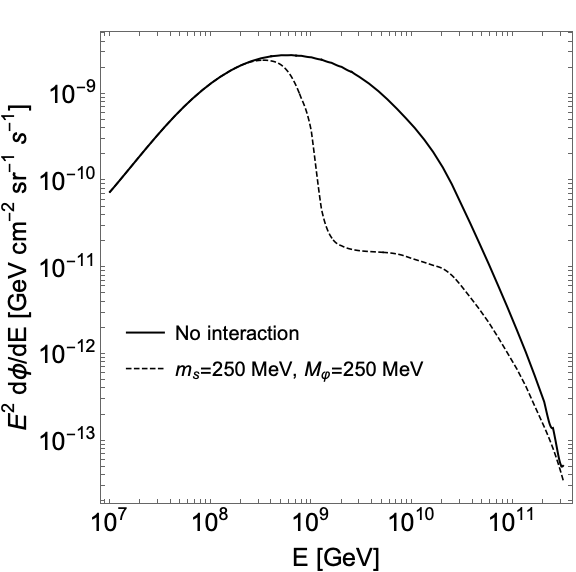}
\caption{Cosmogenic all flavor neutrino power law spectra: the thick line is the flux with no interaction, while the dashed line correspond to the benchmark case described in the text.}\label{plcosmo}
\end{figure}

For the case of cosmogenic fluxes, which are relevant at higher energies, the absorption effect is most important for higher masses of the sterile neutrino and of the scalar mediator. In this part of the parameter space, the constraints from mesons decay are substantially irrelevant, so there is no need to distinguish between the two case studied above. We will therefore analyze as a single choice the case $\lambda_e=\lambda_\mu=\lambda_\tau=1$, $m_s=250$ MeV and $M_\varphi=250$ MeV. In Figure \ref{plcosmo} we show the effect of the interaction on the cosmogenic flux. We can observe that the effect is maximal around $10^{9 \div 10}$ GeV. We address the reader to our previous paper \cite{Fiorillo:2020jvy} for more details also in relation to future experiments. 

\begin{figure}[h!]
\centering
\includegraphics[width=0.5\textwidth]{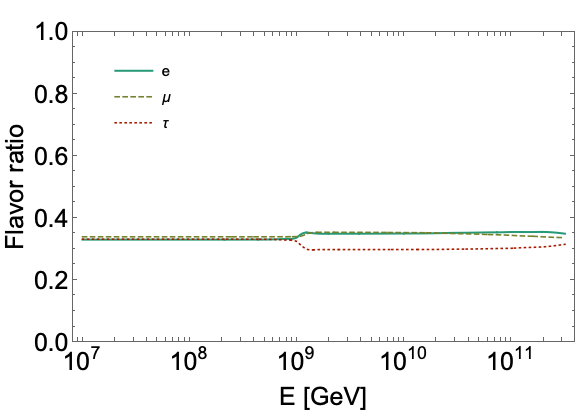}
\caption{Flavor ratio at Earth  as a function of the energy for the cosmogenic benchmark case in the text ($m_s=250$ MeV, $M_\varphi=250$ MeV, $\lambda_e=\lambda_\mu=\lambda_\tau=1$).}\label{frcos}
\end{figure}

As in the astrophysical neutrino case,  also for the cosmogenic flux  we analyze the flavor structure as a function of energy, as shown  in Figures \ref{frcos}.

\section{Conclusions} \label{sec:conclusions}
In this work we have investigated the effects on high- and ultra high- energy active neutrino fluxes due to active-sterile secret interactions mediated by a new pseudoscalar particle. In particular, we  extended our previous  paper \cite{Fiorillo:2020jvy} considering  three active neutrino flavors, leading to an ample freedom of choice for the couplings of the interactions. As a consequence, the laboratory constraints for the meson decays are more relaxed, allowing new regions of the parameter space  otherwise forbidden. Active-sterile neutrino interactions become relevant at very different energy scales depending on the masses of the scalar mediator  and of sterile neutrino. As already found in our previous paper, the final active fluxes can present a measurable depletion observable in future experiments. However, in this study we find that the flux depletion can also occur at lower energy, around the PeV, in the  particular case of only $\nu_\tau- \nu_s$ interaction.  We adopted then two prescriptions for the neutrino flux, namely high-energy represented by a power-law, and ultra high-energy with cosmological origin (cosmogenic) in order to take into account the multi-scale energy phenomenology, varying from $10^6$GeV up to $10^9$GeV.  Remarkably, when the depletion is around $10^6$GeV, this effect could be very interesting for IceCube because it can produce a cutoff-like feature in the spectrum, which could potentially explain  the lack of observed events above 10 PeV.  For larger values of mediator and sterile masses,  the depletion effect  instead could be only visible at larger energy, around $10^9$GeV,  with future experiments like GRAND. Another interesting phenomenological aspect of active-sterile secret interactions is represented by the changing in  the flavor ratio as a function of neutrino energy. This effect could be interesting for next generation of neutrino telescopes like IceCube-Gen2 or KM3NeT. \\

{\it\bf Acknowledgments:} we thank Alessandro Mirizzi for useful discussion.
This work was partially supported by the research grant number 2017W4HA7S ``NAT-NET: Neutrino and Astroparticle Theory Network'' under the program PRIN 2017 funded by the Italian Ministero dell'Universit\`a e della Ricerca (MUR). The authors acknowledge partial support by the research project TAsP (Theoretical Astroparticle Physics) funded by the Instituto Nazionale di Fisica Nucleare (INFN).

\appendix
\section*{Appendices}
\section{Derivation of the transport equations}
We will denote by $F_\alpha (E,\mathbf{n},\mathbf{r})=\frac{d\phi}{dEd\Omega}$ the flux of $\alpha=a,s$ neutrinos passing at point $\mathbf{r}$ with a direction in a small solid angle interval around $\mathbf{n}$. Taking into account the collisions with the CNB neutrinos, and neglecting for the moment the redshifting effects, the transport equations for the active neutrinos will be
\begin{eqnarray}
&& \mathbf{n}\cdot\frac{\partial F_a}{\partial \mathbf{r}}=-n \sigma_a F_a\\ \nonumber  && +\int F_s(E',\mathbf{n}',\mathbf{r}) n\frac{d\sigma_{sa}}{dEd\Omega}dE' d\Omega'+\overline{\rho} f(E,\Omega),
\end{eqnarray}
where the source term is assumed to be originated from a collection of sources, each of which produces $\frac{dN}{dEdtd\Omega}=f(E,\Omega)$ neutrinos per unit time per unit energy interval per unit solid angle. The sources are assumed to be distributed with a number density $\overline{\rho}$. Since neutrinos are highly relativistic, the collisions are strongly forward, with an emission angle suppressed by a factor of $\frac{E_s}{m_s}\sim 10^{-10}$. Under these conditions, we can assume in the integral in the second term that $F_s(E',\mathbf{n}',\mathbf{r})=F_s(E',\mathbf{n},\mathbf{r})$, so that the integration over the solid angle $\Omega'$ can be performed directly, and the equation simplifies to
\begin{equation} \label{eqa2}
\mathbf{n}\cdot\frac{\partial F_a}{\partial \mathbf{r}}=-n \sigma_a F_a+ \int F_s(E',\mathbf{n},\mathbf{r}) n\frac{d\sigma_{sa}}{dE}dE'+\overline{\rho} f(E,\Omega).
\end{equation}
If the sources are considered to be isotropically distributed, then $F_a$ and $F_s$ are isotropic as well, which allows us to simplify Eq. \ref{eqa2} to
\begin{equation}
\frac{\partial F_a}{\partial r}=-n \sigma_a F_a+ \int F_s(E') n\frac{d\sigma_{sa}}{dE}dE'+\overline{\rho} f(E,\Omega).
\end{equation}
Parameterizing the radial distance in terms of the redshift
\begin{equation}
dr=\frac{dz}{H(z)(1+z)},
\end{equation}
and further taking into account that in the free propagation the flux decreases as $F\sim \frac{1}{a^2(t)}$, we obtain
\begin{eqnarray} \label{eqa6}
\nonumber 2H(z) F_a -H(z)(1+z)\frac{\partial F_a}{\partial z}=-n \sigma_a F_a\\  + \int F_s(E') n\frac{d\sigma_{sa}}{dE}dE'+\overline{\rho} f(E,\Omega).
\end{eqnarray}
A final effect to take into account is the redshifting of the energy. This implies that the partial derivative in redshift should be substituted by the transport derivative along the characteristic lines. On such lines the energy changes as $E(1+z)$. The equation then takes the form
\begin{eqnarray}
\nonumber 2H(z) F_a -H(z)(1+z)\left[\frac{\partial F_a}{\partial z}+\frac{\partial F_a}{\partial z} \frac{E}{1+z}\right]=\\  -n \sigma_a F_a + \int F_s(E') n\frac{d\sigma_{sa}}{dE}dE'+\overline{\rho} f(E,\Omega).
\end{eqnarray}
Finally, by expressing $\overline{\rho}=\rho(z)(1+z)^3$, where$\rho(z)$ is the comoving number density and, defining $\Phi_a=\frac{F_a}{(1+z)^2}$, Eq. \ref{eqa6} takes the final form
\begin{eqnarray}
\nonumber &-& H(z)(1+z)\left[\frac{\partial \Phi_a}{\partial z}+\frac{\partial \Phi_a}{\partial z} \frac{E}{1+z}\right]=\\ \nonumber &-&n(z) \sigma_a \Phi_a + \int \Phi_s(E') n(z)\frac{d\sigma_{sa}}{dE}dE' \\ &+&\rho(z)(1+z) f(E,\Omega).
\end{eqnarray}
Notice that, since we need $F_a$ evaluated at Earth, where $z=0$, the variable $\Phi_a$ can be directly used in place of the correct flux $F_a$. In fact, at Earth we have $F_a(z=0)=\Phi_a(z=0)$.
We therefore recover the form of the transport equations given in the text. Analogous passages can be made on the transport equation for the sterile flavor.

\section{Treatment of the sterile neutrino decay}

As mentioned in the text, if $m_s>M_\varphi$ the sterile neutrinos produced after the interaction of the astrophysical active neutrinos with the CNB are not stable and can decay to an active neutrino and a scalar mediator. The lifetime for the decay of a sterile neutrino with energy $E$ into an $i$th active neutrino is
\begin{equation}
\tau_i=\frac{8\pi m_s^2 E}{|\sum_\alpha U_{\alpha i}\lambda_\alpha|^2 (m_s^2-M_\varphi^2)^2},
\end{equation}
where $U_{\alpha i}$ are the elements of the PMNS matrix.

Because of relativistic boosting, for the energies we are interested in, the active neutrinos are produced nearly in the same direction as the original sterile neutrino. Their energy distribution is
\begin{equation}\label{endist}
\frac{dN}{dE_a}(E\to E_a)=\frac{m_s}{m_s^2-M_\varphi^2}\frac{1}{\sqrt{\frac{E^2}{m_s^2}-1}},
\end{equation}
where $E_a$ is the energy of the active neutrino, which can take values between the extrema
\begin{equation}
E_{1,2}=\frac{E(m_s^2-M_\varphi^2)}{2m_s^2}\left[1\pm\sqrt{1-\frac{m_s^2}{E^2}}\right].
\end{equation}
The probability of decaying into the $i$th mass eigenstate is
\begin{equation}
P_i=\frac{|\sum_\alpha U_{\alpha i}\lambda_\alpha|^2}{\sum_i |\sum_\alpha U_{\alpha i}\lambda_\alpha|^2}.
\end{equation}
A simple numerical estimate shows that, for masses of the sterile neutrinos even slightly larger than the scalar masses, the distances over which the sterile neutrinos are expected to decay are much smaller than their mean free path for collision with the CNB. Under these conditions, decays are so fast that one can assume that, as soon as a sterile neutrino is produced, it immediately decays into an active neutrino with the energy distribution determined above. The transport equations in this regime can be therefore approximated by assuming that the flux of sterile neutrinos injected per unit path length by the collisions of active neutrinos with the CNB, namely
\begin{eqnarray}
&&-H(z)(1+z)\frac{\partial \Phi_s(z,E)}{\partial z}=\\ \nonumber &&\sum_i \int dE' \Phi_i(z,E') \frac{d\sigma_{as}}{dE}(E'\to E) n(z),
\end{eqnarray}
is completely converted into an active neutrino flux with the energy distribution in Eq. \eqref{endist}. We can therefore write the equations as
\begin{eqnarray}
&&H(z)(1+z)\left(\frac{\partial \Phi_i (z,E)}{\partial z} + \frac{\partial \Phi_i(z,E)}{\partial E}\frac{E}{1+z}\right)=\nonumber \\  &&n(z) \sigma_i \Phi_i (z,E)\nonumber\\&&- \int dE' \int dE'' \frac{dN}{dE}(E'\to E) P_i \times \nonumber \\ &&\sum_j \Phi_j(z,E'') \frac{d\sigma_{as}}{dE'}(E''\to E') n(z)\nonumber \\&&-\rho(z)(1+z) f(E) \xi_i.
\end{eqnarray}
In order to estimate the relevance of the regeneration term, we have followed the approach in the text and treated it as a perturbation, so that we could verify \textit{a posteriori} whether its corrections can be considered small. We found that, differently from the regime $m_s<M_\varphi$, the first order perturbative corrections from the regeneration term can significantly change the spectrum. Nevertheless, by estimating the second order corrections, we verified that the perturbative results are trustworthy, since the second order corrections are much smaller than the first order ones, thereby ensuring the convergence of the perturbation series.

In the text, where it is not differently specified, we have treated the regime $m_s>M_\varphi$, in which decay is relevant, with the perturbative approach for the regeneration term.

The validity of this perturbative treatment, which is based on the fact that decays happen faster than all the other collision processes, is restricted to values of $m_s$ not too close to $M_\varphi$, since otherwise the decay lifetime might become too long and the decay might become ineffective. However, some simple numerical estimates show that that this caveat is only effective for $m_s$ differing by $M_\varphi$ much less than $1$ MeV. Under these conditions, the usual treatment without the decay term shows that regeneration can simply be neglected.

\end{document}